\documentclass{SciPost}

\hypersetup{
    colorlinks,
    linkcolor={red!50!black},
    citecolor={blue!50!black},
    urlcolor={blue!80!black}
}

\usepackage[bitstream-charter]{mathdesign}
\DeclareSymbolFont{usualmathcal}{OMS}{cmsy}{m}{n}
\DeclareSymbolFontAlphabet{\mathcal}{usualmathcal}

\fancypagestyle{SPstyle}{
\fancyhf{}
\lhead{\colorbox{scipostblue}{\bf \color{white} ~SciPost Physics}}
\rhead{{\bf \color{scipostdeepblue} ~Submission }}

\fancyfoot[C]{\textbf{\thepage}}
}
\usepackage{bm} 

\usepackage{graphicx} 
\usepackage{caption}
\begin{document}

\pagestyle{SPstyle}

\begin{center}{\Large \textbf{\color{scipostdeepblue}{
Transmission matrix measurement of a single Mie scatterer\\
}}}\end{center}

\begin{center}\textbf{
Xiaomeng Sui\textsuperscript{1$\star$}
and Allard Mosk\textsuperscript{1}
}\end{center}
\begin{center}
{\bf 1} Debye Institute for Nanomaterials Science and Center for Extreme Matter and Emergent Phenomena, Utrecht University, Utrecht, the Netherlands
\\[\baselineskip]
$\star$ \href{mailto:email1}{\small x.sui@uu.nl}
\end{center}

\section*{\color{scipostdeepblue}{Abstract}}
\textbf{\boldmath{%
Transmission matrices are valuable tools to describe and control light transport through scattering media. There are only a few cases where the transmission matrix can be compared to microscopic theories. Here we measure the polarization-complete transmission matrix of a single dielectric sphere using off-axis holography with angle scanning and reconstruct complex fields in both transmission and reflection under circular polarization. After aberration correction and angular mapping, the scattering amplitude extracted from the transmission matrix closely follows Mie theory. This work provides a calibrated benchmark for angle-resolved transmission matrix measurement and enables quantitative characterization of spherical and quasi-spherical scatterers.
}}

\vspace{\baselineskip}

\noindent\textcolor{white!90!black}{%
\fbox{\parbox{0.975\linewidth}{%
\textcolor{white!40!black}{\begin{tabular}{lr}%
  \begin{minipage}{0.6\textwidth}%
    {\small Copyright attribution to authors. \newline
    This work is a submission to SciPost Physics. \newline
    License information to appear upon publication. \newline
    Publication information to appear upon publication.}
  \end{minipage} & \begin{minipage}{0.4\textwidth}
    {\small Received Date \newline Accepted Date \newline Published Date}%
  \end{minipage}
\end{tabular}}
}}
}


\vspace{10pt}
\noindent\rule{\textwidth}{1pt}
\tableofcontents
\noindent\rule{\textwidth}{1pt}
\vspace{10pt}


\section{Introduction}
Transmission matrices (TMs) provide a linear description of light transport through scattering systems by explicitly mapping a set of input optical modes to corresponding output modes \cite{popoff2010measuring}. This framework has enabled powerful approaches for understanding and controlling wave propagation in complex media, such as wavefront shaping and light focusing through strongly scattering samples \cite{mosk2012controlling}, by identifying highly transmitting eigenchannels \cite{dorokhov1984coexistence, vellekoop2008universal, yu2013measuring, popoff2014coherent, yilmaz2019transverse}. Beyond wavefront control, transmission matrices provide direct access to wave-transport phenomena such as angular memory effects \cite{yilmaz2019angular}, invariance property \cite{pierrat2014invariance, pai2021scattering}, symmetry-induced anti-reflection \cite{horodynski2022anti}, and Anderson localization \cite{abrahams1979scaling}. Most existing TM studies address highly disordered media \cite{cao2022shaping}, where the scattering operator is not predictable for a given sample. As a result, TM interpretation is often statistical, relying on mesoscopic light transport \cite{rotter2017light} and random-matrix theory \cite{beenakker1997random}, and nowadays also includes Fisher information \cite{bouchet2021maximum} and matrix fingerprints \cite{le2025detection}. While this statistical viewpoint has been extremely successful, it does not readily allow the validation of a field operator against an analytically solvable three-dimensional scatterer. By contrast, measuring the TM of a single, well-defined scatterer provides a complementary situation where direct, model-based parameter inference becomes possible.

An isolated dielectric microsphere provides a well-controlled test case because its scattering problem admits an exact analytical solution. For a homogeneous spherical particle illuminated by a monochromatic plane wave, Maxwell’s equations can be solved by expanding both the incident and scattered fields in a complete basis of vector spherical harmonics \cite{lorenz1890lysbevaegelsen, mie1908beitrage}. In this framework, the scattering response is fully determined by a discrete set of scattering coefficients, which describe how each spherical harmonic mode is excited and radiated by the particle. Mie theory provides this solution and expresses the far-field response through the amplitude scattering matrix and the associated scattering amplitude functions, which account for the combined contributions of electric and magnetic multipoles \cite{bohren2008absorption}. Because this description is exact and parameterized only by the size parameter and the refractive index, light scattering from a single spherical particle is highly predictable. In optical trapping, Mie theory is used to describe radiation-pressure forces and to calibrate trap dynamics \cite{pesce2020optical, juan2011plasmon}. In sensing and metrology, the angle-dependent and wavelength-dependent Mie features provide a sensitive framework for estimating particle size, refractive index, and the surrounding environment \cite{altman2021holographic, teraoka2003perturbation, zijlstra2007spatial}. In microscopy and holographic imaging, Mie theory provides quantitative forward models for elastic scattering contrast \cite{lee2007characterizing, sersic2011fourier}. It has also been used to analyze plasmonic metal spheres as nanoantennas \cite{langguth2016exact} in fluorescence correlation spectroscopy.

In this article, we measure the polarization-complete transmission matrix of a single spherical dielectric microparticle using off-axis holography with uniform sampling of illumination angles across the pupil, which is performed in a circular polarization basis. From the measured incident and scattered fields, an angle-resolved transmission matrix is formed, and the angle-dependent scattering amplitude is extracted. The co-polarized scattering amplitude is fitted with Mie theory to characterize the sphere parameters, which are then used to verify the cross-polarized channel without additional fitting. The fitted forward model is further tested in reflection geometry, where the measured back-scattered fields are influenced by coherent multiple scattering effects at the air-glass boundary. Including these interface-related contributions enables a consistent interpretation of the reflection data beyond an ideal free-space scattering model.

\section{Mie scattering}
\label{sec:mie_scattering}
In our theoretical description, we consider a deterministic and well-defined scattering system: a single, homogeneous spherical particle with a given refractive index $n_p$ and radius $a_r$, in a medium of refractive index $n_m$. 

\subsection{Amplitude scattering matrix}
\label{subsec:mie_theory}

\begin{figure}[t]
    \centering
    \includegraphics[width=\linewidth]{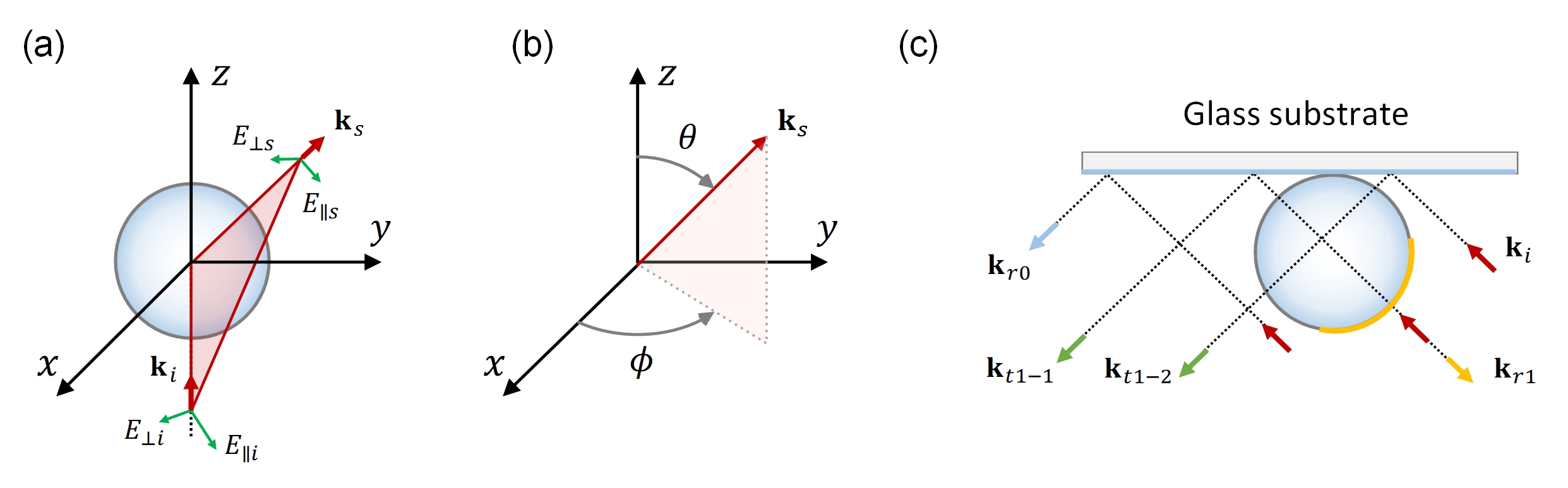}
    \caption{
        \textbf{Scattering by a single sphere.} (a)  Scattering plane spanned by the incident wave vector and the scattering wave vector. (b) Scattering (polar) angle $\theta$ and azimuthal angle $\phi$ in spherical coordinates. (c) Sphere on a glass substrate. In reflection, the measured field includes additional multiple reflections between the sphere and the air-glass interface.
    }
    \label{fig:figure1}
\end{figure}

The scattering geometry is specified by an incident monochromatic wave vector $\mathbf{k}_i$ and a scattered wave vector $\mathbf{k}_s$. The incident and scattered fields are decomposed into components parallel $(\parallel)$ and perpendicular $(\perp)$ to the scattering plane spanned by the unit vectors $\hat{\mathbf{k}}_i$ and $\hat{\mathbf{k}}_s$, as depicted in Fig. \ref{fig:figure1}(a). The observation direction is parameterized by the scattering angle $\theta$ and the azimuthal angle $\phi$ in spherical coordinates $(r,\theta,\phi)$, as shown in Fig. \ref{fig:figure1}(b). For the convenience of defining the scattering angles and polarization basis, the coordinate system is chosen such that $\hat{\mathbf{k}}_i=\hat{\mathbf{z}}$. This is a coordinate choice and does not restrict generality because a homogeneous sphere is rotationally symmetric.

In the far field, the relation between the incident and scattered electric-field components is written in terms of the amplitude scattering matrix \cite{bohren2008absorption}
\begin{equation}
\begin{pmatrix}
E_{\parallel s} \\
E_{\perp s}
\end{pmatrix}
=
\frac{e^{ik(r-z)}}{-ikr}
\begin{pmatrix}
S_{1} & S_{3} \\
S_{4} & S_{2}
\end{pmatrix}
\begin{pmatrix}
E_{\parallel i} \\
E_{\perp i}
\end{pmatrix}\;,
\label{eq:amp_scattering_matrix}
\end{equation}
where the exponential prefactor represents spherical-wave propagation, and the matrix elements $S_{1}$, $S_{2}$, $S_{3}$, and $S_{4}$ are angle-dependent scattering amplitudes. For a homogeneous sphere, rotational symmetry removes any dependence on $\phi$ and eliminates polarization-mixing terms, so that $S_{3}=S_{4}=0$ and the response is fully characterized by $S_1(\theta)$ and $S_2(\theta)$.

The functions $S_1$ and $S_2$ are expressed as convergent series in terms of the Mie coefficients and the angular functions, the explicit expressions of which are given in Appendix~\ref{appendix:mie_theory}. The Mie coefficients depend on the size parameter $x$ and the relative refractive index $m$, given as
\begin{equation}
x = k_m a_r = \frac{2\pi n_m a_r}{\lambda_0}\;,
\qquad
m=\frac{n_p}{n_m}\;,
\label{eq:size_parameter}
\end{equation}
where $\lambda_0$ is the vacuum wavelength and $k_m$ is the wavenumber in the surrounding medium. Using the amplitude scattering matrix in Eq.~\eqref{eq:amp_scattering_matrix}, the scattered field corresponding to any incident direction can be computed. In our simulations, the transmission matrix is evaluated at the angle space, so the common propagation prefactor is omitted, and the scattered field is evaluated directly on the detection plane. 

For the numerical implementation, we developed a Python code following the standard Mie-theory formulation in \cite{bohren2008absorption}. The complex refractive index and the particle size parameter are used as inputs to compute the Mie coefficients and the corresponding scattering amplitudes $S_{1}$ and $S_{2}$. The implementation was validated against \texttt{miepython}~\cite{prahl_miepython_2025}.

\subsection{Circular polarization}
\label{subsec:cir-polarization}
So far, the scattering relation has been expressed in the linear polarization basis defined by the scattering plane. To preserve the rotational symmetry of the sphere, it is convenient to work in the circular polarization basis, where the field is decomposed into left-handed ($L$) and right-handed ($R$) circular components, denoted by $(E_L, E_R)^{\mathsf{T}}$. We introduce a unitary change-of-basis matrix that relates the linear and circular polarization bases:
\begin{equation}
\mathbf{R}_\mathrm{cir}=\frac{\sqrt{2}}{2}
\begin{pmatrix}
1 & 1\\
-i & i
\end{pmatrix}\;,
\label{eq:R_def}
\end{equation}
such that the linear polarization components are obtained from the circular polarization components via
\begin{equation}
\begin{pmatrix}
E_{\parallel s} \\
E_{\perp s}
\end{pmatrix}
=
\mathbf{R}_\mathrm{cir}
\begin{pmatrix}
E_{L s} \\
E_{R s}
\end{pmatrix}\;,
\qquad
\begin{pmatrix}
E_{\parallel i} \\
E_{\perp i}
\end{pmatrix}
=
\mathbf{R}_\mathrm{cir}
\begin{pmatrix}
E_{L i} \\
E_{R i}
\end{pmatrix}\;.
\label{eq:lin_circ_inc_scat}
\end{equation}
Substituting Eq.~\eqref{eq:lin_circ_inc_scat} into Eq.~\eqref{eq:amp_scattering_matrix}, we obtain the scattering relation in the circular polarization basis:
\begin{equation}
\begin{pmatrix}
E_{L s} \\
E_{R s}
\end{pmatrix}
=
\frac{e^{ik(r-z)}}{-ikr}\,
\mathbf{R}_\mathrm{cir}^{-1}
\begin{pmatrix}
S_2 & 0 \\
0   & S_1
\end{pmatrix}
\mathbf{R}_\mathrm{cir}
\begin{pmatrix}
E_{L i} \\
E_{R i}
\end{pmatrix}\;.
\label{eq:circular_scattering_general}
\end{equation}
Evaluating the similarity transform gives
\begin{equation}
\mathbf{R}_\mathrm{cir}^{-1}
\begin{pmatrix}
S_2 & 0 \\
0   & S_1
\end{pmatrix}
\mathbf{R}_\mathrm{cir}
=
\frac{1}{2}
\begin{pmatrix}
S_1+S_2 & S_2-S_1\\
S_2-S_1 & S_1+S_2
\end{pmatrix}\;,
\label{eq:similarity_result}
\end{equation}
so that
\begin{align}
E_{L s}
&=
\frac{e^{ik(r-z)}}{-ikr}\,
\frac{1}{2}\Big[(S_1+S_2)\,E_{L i}+(S_2-S_1)\,E_{R i}\Big]\;,
\label{eq:ELs_general}\\
E_{R s}
&=
\frac{e^{ik(r-z)}}{-ikr}\,
\frac{1}{2}\Big[(S_2-S_1)\,E_{L i}+(S_1+S_2)\,E_{R i}\Big]\;.
\label{eq:ERs_general}
\end{align}
In the circular polarization basis, the incident and scattered field vectors are then written as
$\mathbf{E}_{LRi}=(E_{Li},E_{Ri})^{\mathsf{T}}$ and $\mathbf{E}_{LRs}=(E_{Ls},E_{Rs})^{\mathsf{T}}$. The sphere response is represented by the amplitude scattering matrix $\mathbf{S}_{LR}$ derived from Eqs.~(\ref{eq:ELs_general}) and (\ref{eq:ERs_general}), which yields:
\begin{equation}
\mathbf{E}_{LRs}=\mathbf{S}_{LR}\,\mathbf{E}_{LRi}.
\label{eq:Es_Ein}
\end{equation}
For an incident field that is purely left-circularly polarized, which means $E_{R i}=0$, Eqs.~\eqref{eq:ELs_general} and \eqref{eq:ERs_general} reduce to
\begin{align}
E_{L s}
&=
\frac{e^{ik(r-z)}}{-ikr}\,
\frac{1}{2}(S_1+S_2)\,E_{L i}\;,
\label{eq:ELs_Linc}\\
E_{R s}
&=
\frac{e^{ik(r-z)}}{-ikr}\,
\frac{1}{2}(S_2-S_1)\,E_{L i}\;.
\label{eq:ERs_Linc}
\end{align}
These expressions show that, after transforming to the circular polarization basis, the co-polarized component is proportional to $(S_1+S_2)/2$, while the cross-polarized component is proportional to $(S_2-S_1)/2$.

In transmission geometry, a path undergoing multiple reflections between the sphere and the air-glass interface involves at least one Fresnel reflection and one particle reflection. These higher-order multiple scattering are negligible because the glass-reflection and backward-scattering contributions are very small compared with forward scattering. As a result, the transmission field can be well approximated by the sphere scattering followed by transmission through the glass substrate:
\begin{equation}
\mathbf{E}_{LRt}\approx \mathbf{t}\,\mathbf{E}_{LRs},
\label{eq:Et_trans_approx}
\end{equation}
where $\mathbf{t}$ accounts for the transmittance and the propagation. This approximation, which works well for glass and silica, does not generalize to high refractive index materials such as TiO$_2$ or Si \cite{kuznetsov2016optically}.

\subsection{Reflection matrix}
\label{subsec:reflection}
In reflection geometry, the measured field contains all the components scattered back toward the illumination side. Because the sphere is mounted on a planar glass substrate in the experiment, the reflected field contains not only free-space Mie scattering but also additional contributions arising from multiple reflections between the sphere and the air-glass interface, as illustrated in Fig. \ref{fig:figure1}(c). An incident plane wave with wave vector $\mathbf{k}_i$ illuminates the particle under oblique incidence. In addition to the direct scattering into an observation direction $\mathbf{k}_{r1}$, the presence of the substrate introduces a specularly reflected field. This reflected wave can re-illuminate the particle, producing additional effective incident directions ($\mathbf{k}_{t1-1}$ and $\mathbf{k}_{t1-2}$ in the scheme) and additional scattered components. The measured field can therefore be interpreted as a superposition of scattering contributions driven by the direct incident beam and by one or more sphere-glass reflections.

Following the definition of the incident field vectors $\mathbf{E}_{LRi}$ and the amplitude scattering matrix $\mathbf{S}_{LR}$ given in Eq. (\ref{eq:Es_Ein}), the total reflected field is modeled as the coherent sum over multiple re-illumination cycles between the sphere and the air-glass interface: 
\begin{equation}
\mathbf{E}_{LRr}
=
\sum_{n=0}^{N}
\left(\mathbf{S}_{LR}\,\mathbf{\alpha}\right)^{n}\,
\mathbf{S}_{LR}\,\mathbf{E}_{LRi}
\;+\;
\sum_{n=0}^{N}
\left(\mathbf{\alpha}\,\mathbf{S}_{LR}\right)^{n}\,\mathbf{\alpha}\,\mathbf{E}_{LRi}\;,
\label{eq:multipass_two_sums}
\end{equation}
where the first series represents a path first scattered by the sphere, and the second series represents a path first scattered by the air-glass interface. Here $\mathbf{\alpha}$ is a complex operator that accounts for Fresnel reflection \cite{fowles2012introduction} and the phase accumulated in the air gap
\begin{equation}
\mathbf{\alpha}=\mathbf{J}_{LR}\,\exp\!\left(i k_m L\right)\;,
\label{eq:alpha_matrix}
\end{equation}
where $k_m$ is the wavenumber in air, $L$ denotes the effective optical path length between the sphere and the interface, and $\mathbf{J}_{LR}$ is the Fresnel reflection Jones operator in the $(L,R)$ basis,
\begin{equation}
\mathbf{J}_{LR}=\mathbf{R}_\mathrm{cir}^{-1}\begin{pmatrix}
r_p & 0\\
0 & r_s
\end{pmatrix}\mathbf{R}_\mathrm{cir}
=
\frac12
\begin{pmatrix}
r_p+r_s & r_p-r_s\\
r_p-r_s & r_p+r_s
\end{pmatrix}\;.
\label{eq:Fresnel_J_LR}
\end{equation}
In practice, the reflection matrices in Eq. (\ref{eq:multipass_two_sums}) should take into account evanescent as well as propagating waves. The total observed scattering consists of the coherent sum over the direct illumination ($n=0$) and a finite number of sphere-glass reflections ($n\ge 1$), each followed by the corresponding one or more Mie scattering, Fresnel reflection, and phase delay operators. Expanding the equation yields the first few terms of each series:
\begin{align}
\mathbf{E}_{LRr}
=&\;
\Big[
\mathbf{S}_{LR}
+(\mathbf{S}_{LR}\mathbf{\alpha})\mathbf{S}_{LR}
+(\mathbf{S}_{LR}\mathbf{\alpha})^{2}\mathbf{S}_{LR}
+\cdots
\Big]\mathbf{E}_{LRi}
\nonumber\\
&\;+\Big[
\mathbf{\alpha}
+(\mathbf{\alpha}\mathbf{S}_{LR})\mathbf{\alpha}
+(\mathbf{\alpha}\mathbf{S}_{LR})^{2}\mathbf{\alpha}
+\cdots
\Big]\mathbf{E}_{LRi}\;.
\label{eq:two_sums_expanded}
\end{align}
Ignoring higher orders resulting from two or more reflection events from the sphere and the air-glass interface, it is sufficient to retain the five dominant contributions, yielding
\begin{align}
\mathbf{E}_{LRr}\approx\;
&\mathbf{S}_{LR}\,\mathbf{E}_{LRi}
+\mathbf{\alpha}\,\mathbf{E}_{LRi}
+\mathbf{\alpha}\,\mathbf{S}_{LR}\,\mathbf{E}_{LRi}
+\mathbf{S}_{LR}\,\mathbf{\alpha}\,\mathbf{E}_{LRi}
+\mathbf{S}_{LR}\,\mathbf{\alpha}\,\mathbf{S}_{LR}\,\mathbf{E}_{LRi}\;.
\label{eq:four_terms_from_two_sums}
\end{align}
In this approximation, the first term represents the direct sphere contribution (MieBw), and the second term corresponds to the specular reflection from the air-glass interface (G). The third and fourth terms represent the two lowest orders involving both the sphere and the substrate: the third term represents a path of sphere scattering followed by interface reflection (MieFw-G), while the fourth term describes the reverse sequence (G-MieFw). The fifth term describes a sphere-glass-sphere sequence: the field is first scattered by the sphere, then undergoes a substrate reflection, and is scattered again by the sphere (MieFw-G-MieFw). We use Eq. (\ref{eq:four_terms_from_two_sums}), restricted to propagating modes only, to approximate the back-scattered field.

\section{Optical configuration}
\label{sec:another}
The sample used in the experiment is a homogeneous silica (SiO$_2$) microparticle. The image of a single particle under scanning electron microscopy (SEM) is shown in Fig. \ref{fig:figure2}(b) in  Appendix \ref{appendix:Off-axis holography}. This particle has an approximate diameter of 8~\textmu m and a refractive index of $\sim$1.45 (supplier Bangs Laboratories, Fisher, USA). The silica microparticle is almost a perfect sphere with a homogeneous refractive index and therefore well described by Mie theory. 

In the experiment, the complex scattered field is retrieved using off-axis holography, which provides single-shot access to the amplitude and phase of the object field via its interference with a tilted reference beam. The particle is illuminated and observed with high-NA objectives, using a condenser objective (60$\times$, NA~0.98) for illumination and an imaging objective (63$\times$, NA~1.42) for collection. A plane-wave illumination is prepared in a left-handed circular polarization state, and the detected field is decomposed into co-polarized and cross-polarized components. By steering the illumination angle, holograms are recorded for a sequence of incident wave vectors in both transmission and reflection detection channels. Further experimental details are provided in Appendix~\ref{appendix:Off-axis holography}.

\subsection{Pupil aberration removal of high-NA objective}
\label{sec:pupil_model}

In a high-NA microscope, the relation between pupil position and propagation direction cannot be described by the small-angle approximation. High-NA objectives are typically designed to satisfy the Abbe-sine condition \cite{abbe1873beitrage}, so the pupil radius is proportional to the sine of the propagation angle rather than to the propagation angle itself. However, this ideal relation is not perfectly satisfied in practice, because fabrication errors, inhomogeneous refractive index, and optical misalignment introduce a spatially varying phase across the pupil. This pupil aberration distorts the recovered complex fields. Therefore, pupil aberration removal for the imaging objective is performed by estimating the pupil-phase term and subtracting it from the reconstructed complex fields.

\begin{figure}[t]
    \centering
    \includegraphics[width=\linewidth]{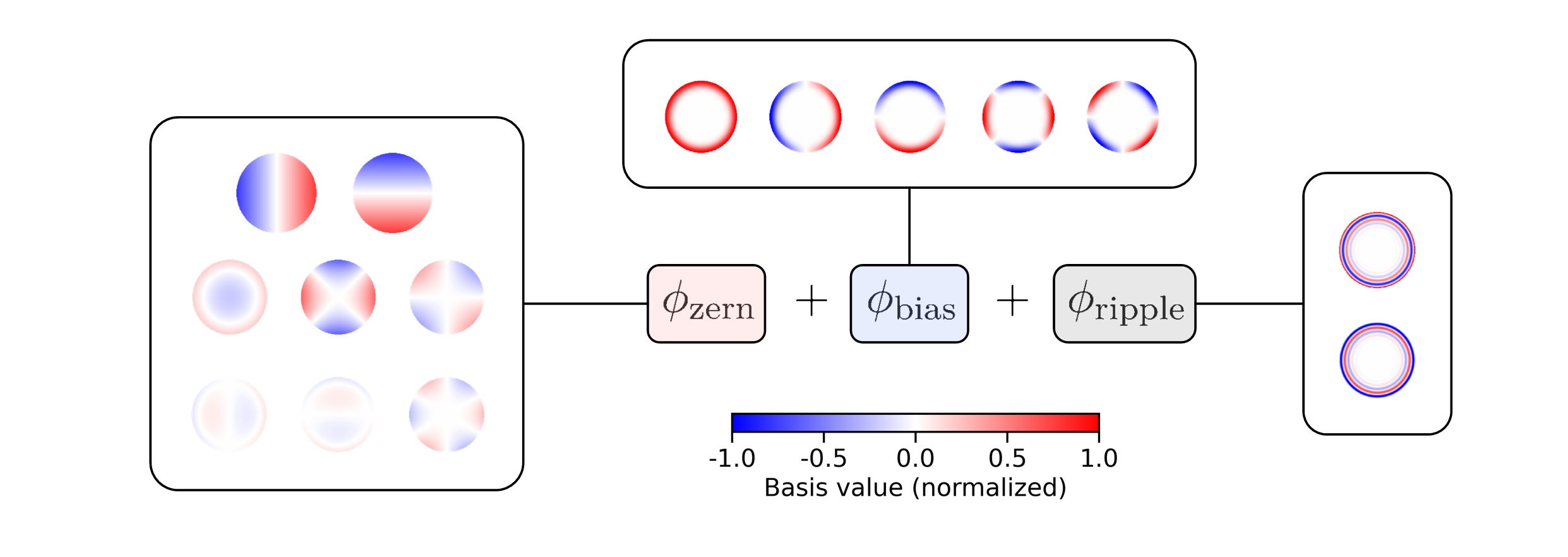}
    \caption{
        \textbf{Pupil-phase aberration model with representative Zernike, edge-bias, and edge-ripple basis functions on the unit pupil.}
    }
    \label{fig:figure3}
\end{figure}

We approximate the pupil aberration by combining a low-order Zernike contribution with two supplementary terms that describe edge features observed in experimentally retrieved pupil phases:
\begin{equation}
\phi_{\mathrm{model}}(\rho,\vartheta)
=
\phi_{\mathrm{zern}}(\rho,\vartheta)
+
\phi_{\mathrm{bias}}(\rho,\vartheta)
+
\phi_{\mathrm{ripple}}(\rho,\vartheta)\;,
\label{eq:phi_model_sum}
\end{equation}
where $(\rho,\vartheta)$ are normalized polar coordinates on the pupil disk, with $\rho \in [0,1]$ and $\vartheta\in[0,2\pi)$. Here $\phi_{\mathrm{zern}}$ is a Zernike expansion on the unit disk. The term $\phi_{\mathrm{bias}}$ models slowly varying angular deviations near the aperture boundary through an angular Fourier series, and $\phi_{\mathrm{ripple}}$ accounts for ring-like oscillations near the boundary through radial sine and cosine components. The corresponding low-order basis functions are shown in Fig.~\ref{fig:figure3}.

The Zernike term is the dominant aberration that varies smoothly across the aperture and is well described by low-order terms of the Zernike polynomial \cite{born2013principles} in polar coordinates
\begin{equation}
\phi_{\mathrm{zern}}(\rho,\vartheta)=\sum_{j=1}^{J} a_j\,Z_j(\rho,\vartheta)\;,
\label{eq:zern_expansion}
\end{equation}
where \(Z_j\) are Zernike modes and \(a_j\) are real-valued coefficients in radians. This basis provides a compact description of low-order wavefront distortions generated by the objective and other optics in the setup. In practice, we truncate the Zernike polynomial to $J=8$, which already captures the major low-order aberrations existing in the experiment.

While Zernike modes efficiently describe global wavefront curvature, experimentally retrieved pupil phases still exhibit additional structure concentrated near the aperture boundary. Such boundary features can arise from a slight optical mismatch of the pupil geometry together with the frequency-domain filtering in the reconstruction of off-axis holography. To model these contributions, we introduce the following edge-localization envelope,
\begin{equation}
g(\rho)=\exp\!\left[-\frac{(\rho-1)^2}{2\sigma^2}\right]\;,
\label{eq:edge_envelope}
\end{equation}
where \(\sigma\) defines the effective radial width of the edge region. The envelope makes the two edge terms negligible over the inner pupil and confines them to a narrow band approaching $\rho = 1$.

The first edge term accounts for slowly varying angular deviations near the pupil boundary, which can arise from optical mismatch and slight fabrication imperfections, and is represented by a low-order angular Fourier series \cite{goodman1969introduction}:
\begin{equation}
\phi_{\mathrm{bias}}(\rho,\vartheta)
=
g(\rho)\left[
b_0+\sum_{m=1}^{M}
\left(
b^{(c)}_m\cos(m\vartheta)+b^{(s)}_m\sin(m\vartheta)
\right)
\right]\;.
\label{eq:edge_bias}
\end{equation}
Here $b_0$, $b^{(c)}_m$, and $b^{(s)}_m$ are real coefficients. The constant term $b_0$ represents an approximately uniform phase offset near the edge. The angular order is limited to $M=5$ to reduce overfitting and to avoid absorbing the global aberrations already captured by $\phi_{\mathrm{zern}}$.

After removing the smooth wavefront component, ring-like oscillations are also observed around the pupil boundary. We model this behavior using a radial sine and cosine function in $\rho$, weighted by the same edge envelope \cite{noll1979effect}:
\begin{equation}
\phi_{\mathrm{ripple}}(\rho,\vartheta)
=
g(\rho)\left[
c\cos\left(2\pi f\rho\right)+d\sin\left(2\pi f\rho\right)
\right]\;.
\label{eq:ripple_term}
\end{equation}
The coefficients $c$ and $d$ are real values, and $f$ denotes the ripple frequency in cycles. This form keeps the model linear in the coefficients while allowing the ripple to vary. The ripple frequency is fixed to $f=12$ based on our calibration, and the remaining coefficients are estimated by minimizing the fitting error.

Given fixed hyperparameters $\sigma$, $M$, and $f$, the model is linear in the coefficient vector
\begin{equation}
\boldsymbol{\beta} =
\left[
a_1,\ldots,a_J,\,
b_0,b^{(c)}_1,b^{(s)}_1,\ldots,b^{(c)}_M,b^{(s)}_M,\,
c,d
\right]^{\top}\;.
\label{eq:beta_def}
\end{equation}
The coefficients are estimated over the valid pupil region $\Omega_M$, so that the fit is implemented to the inner aperture in $k$-space. Let the measured phase be $\phi_{\mathrm{meas}}$, then the parameters are obtained by minimizing a weighted least-squares objective
\begin{equation}
\boldsymbol{\beta}^\star
=
\arg\min_{\boldsymbol{\beta}}
\sum_{(\rho,\vartheta)\in\Omega_M}
\left[
\phi_{\mathrm{meas}}-
\phi_{\mathrm{model}}(\boldsymbol{\beta})
\right]^2.
\label{eq:masked_ls}
\end{equation}
This optimization can be solved as a linear least-squares problem \cite{golub1971singular}. After estimating $\boldsymbol{\beta}^\star$, the fitted aberration phase $\phi_{\mathrm{model}}$ is evaluated on the pupil grid and subtracted from the measured pupil phase to obtain an aberration-compensated pupil phase for subsequent reconstruction.

\begin{figure}[t]
    \centering
    \includegraphics[width=\linewidth]{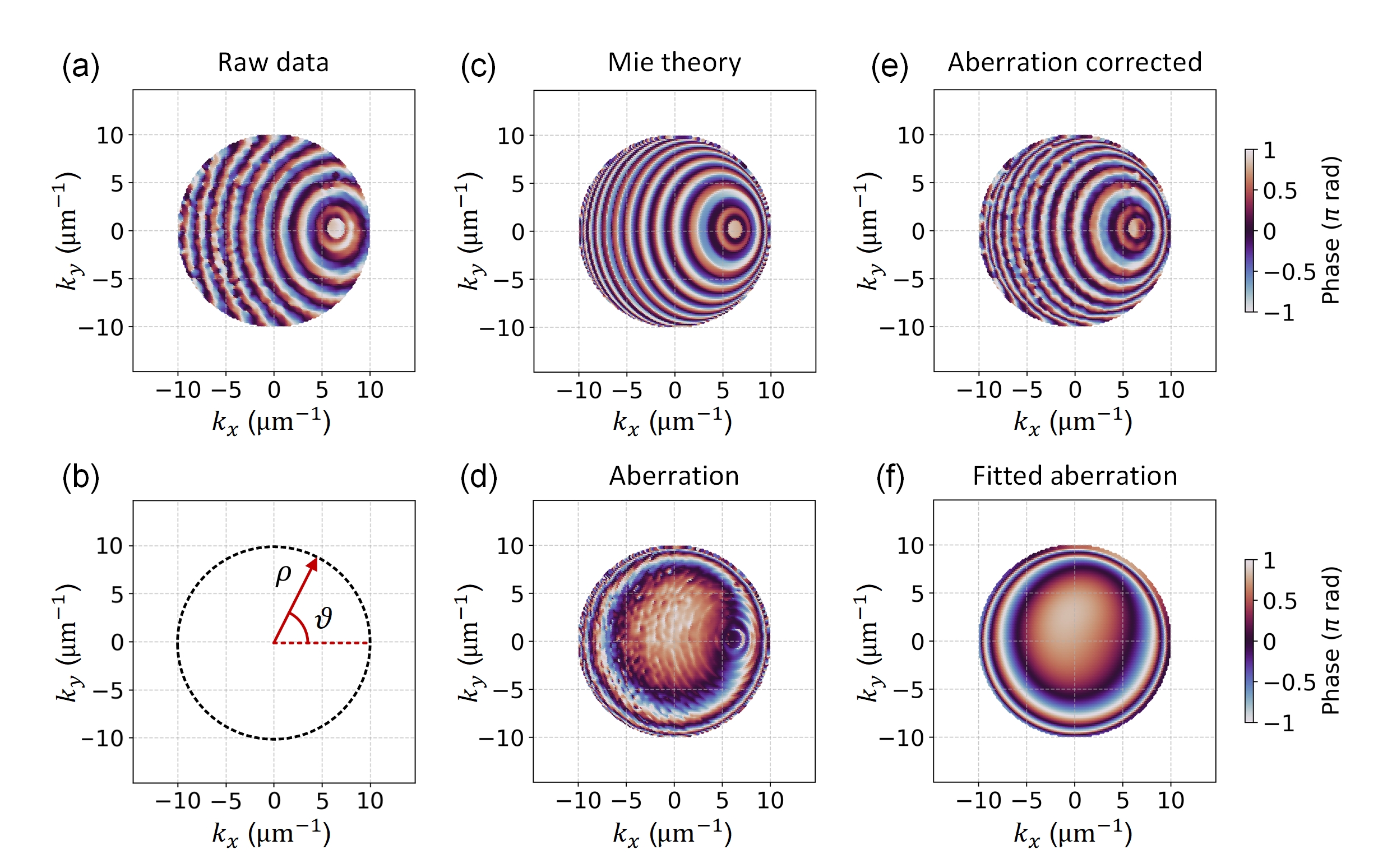}
    \caption{
        \textbf{Pupil aberration estimation and correction of the measurement on the spherical particle.} (a) Measured pupil phase of the co-polarized scattered field from a spherical particle for a single incident vector $\mathbf{k}_i=(6.361,\, 0.326,\, 7.620)$~\textmu$\mathrm{m}^{-1}$. (b) Normalized polar coordinates inside the pupil disk. (c) Theoretical Mie phase for comparison. (d) Estimated pupil aberration. (e) Phase after aberration removal. (f) Fitted aberration from the model.
    }
    \label{fig:figure4}
\end{figure}

The calibration and correction procedure for the pupil phase from the forward-scattered field is thereby summarized in Fig. \ref{fig:figure4}. The experimentally measured pupil-plane phase of a spherical particle under single-angle illumination is shown in Fig.~\ref{fig:figure4}(a) in the $(k_x,k_y)$ pupil coordinates. Within the pupil disk, the normalized polar coordinates are established to parameterize the aberration, as illustrated in Fig.~\ref{fig:figure4}(b). Applying the same incident wave vector, the theoretical Mie phase on this pupil plane is simulated and shown in Fig.~\ref{fig:figure4}(c). The pupil aberration is estimated by comparing the experimentally measured phase with the expected phase response of the spherical particle, which is shown in Fig. \ref{fig:figure4}(d). With the optimization, the aberration is obtained using our aberration model in Eq. (\ref{eq:phi_model_sum}). Subtracting the fitted aberration from the measured pupil phase yields the corrected pupil phase, shown in Fig.~\ref{fig:figure4}(e), and the fitted aberration is shown in Fig.~\ref{fig:figure4}(f). After correction, the measured phase becomes more consistent with this prediction. Cross-validation across multiple illumination angles confirms that our aberration model captures a systematic, pupil-dependent phase distortion.

\begin{figure}[t]
    \centering
    \includegraphics[width=\linewidth]{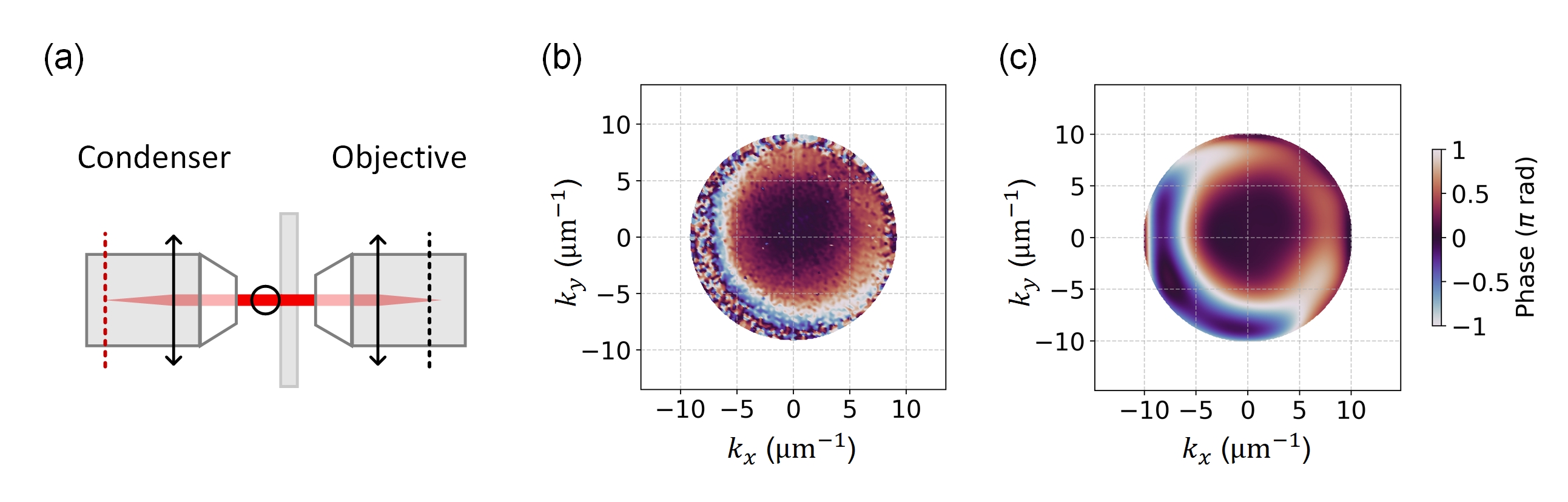}
    \caption{
        \textbf{Condenser pupil aberration estimation.} (a) Condenser and imaging objective in the measurement. (b) Sampled map of the condenser pupil phase obtained through illumination scanning. (c) Fitted condenser pupil aberration predicted by the model. 
    }
    \label{fig:figure5}
\end{figure}

In addition to the imaging objective, the pupil aberration associated with the condenser objective is estimated. This phase aberration mainly influences the back-scattered field, as illustrated in Fig.~\ref{fig:figure5}(a). For each illumination angle, we compute a single representative phase value by averaging the fitted phase over the central pupil region with $\rho\le0.5$. 
Using more than 3000 illumination angles, the phase values obtained for each angle form a sampled map of the condenser pupil aberration, shown in Fig.~\ref{fig:figure5}(b). 
The condenser aberration is then fitted using the same optimization framework as in Eq. (\ref{eq:masked_ls}), and the resulting model prediction is shown in Fig.~\ref{fig:figure5}(c).

\subsection{Angle-resolved transmission matrix}
\label{sec:angle-space}

After off-axis holography reconstruction and pupil aberration removal, each illumination provides a complex field represented in $k$-space. With a single illumination, let $\mathbf{E}\in\mathbb{C}^{N\times 1}$ denote the measured complex values sampled at a series of wave vectors $\{\mathbf{k}_n\}_{n=1}^{N}$, and let $\mathbf{E}_\mathrm{glass}\in\mathbb{C}^{N\times 1}$ denote the corresponding incident-field values obtained from a reference measurement under the same $k$-space sampling.
The measured field contains an unscattered part and a scattered part:
\begin{equation}
\mathbf{E}\approx \gamma\,\mathbf{E}_\mathrm{glass}+\mathbf{E}_s\;,
\end{equation}
where $\gamma \approx e^{i\varphi}$ is a complex correction factor that compensates for possible thickness variations between the reference position and the particle position. The factor $\gamma$ is chosen to make $\gamma\,\mathbf{E}_\mathrm{glass}$ match the unscattered part of $\mathbf{E}$ as closely as possible. It is given by the Moore-Penrose pseudoinverse \cite{penrose1955generalized, moore1920reciprocal} with a threshold parameter of $2.25 \times 10^{-3}$,
\begin{equation}
\gamma = \mathbf{E}_\mathrm{glass}^{+}\mathbf{E}
= \big(\mathbf{E}_\mathrm{glass}^{\dagger}\mathbf{E}_\mathrm{glass}\big)^{-1}\mathbf{E}_\mathrm{glass}^{\dagger}\mathbf{E}\;,
\end{equation}
and the scattered field is obtained by subtracting this matched unscattered contribution,
\begin{equation}
\mathbf{E}_s = \mathbf{E} - \gamma\,\mathbf{E}_\mathrm{glass}
= \Big(\mathbf{I}-\mathbf{E}_\mathrm{glass}\mathbf{E}_\mathrm{glass}^{+}\Big)\mathbf{E}\;.
\end{equation}
Here $(\cdot)^{+}$ denotes the pseudoinverse and $(\cdot)^{\dagger}$ denotes the Hermitian transpose.  Assuming a focus on the $z=0$ plane of the particle coordinates, the scattering amplitude values $\mathbf{A}\in\mathbb{C}^{N\times 1}$ are obtained by normalizing the scattered field with the incident field at each sampled wave vector:
\begin{equation}
A_n=\frac{(E_s)_n}{(E_i)_n},\qquad n=1,\ldots,N\;.
\end{equation}
Each scattering amplitude value is associated with the corresponding wave vector. For Mie scattering, the wave number in the medium is constant, so the values lie on the surface $|\mathbf{k}|=k_m$, forming a spherical shell in $k$-space. However, the scattering amplitude needs to be expressed as a function of the scattering angle $\theta$ using the spherical coordinates defined in Section \ref{subsec:mie_theory}, so that data acquired under different illumination angles can be compared and combined.

\begin{figure}[t]
    \centering
    \includegraphics[width=\linewidth]{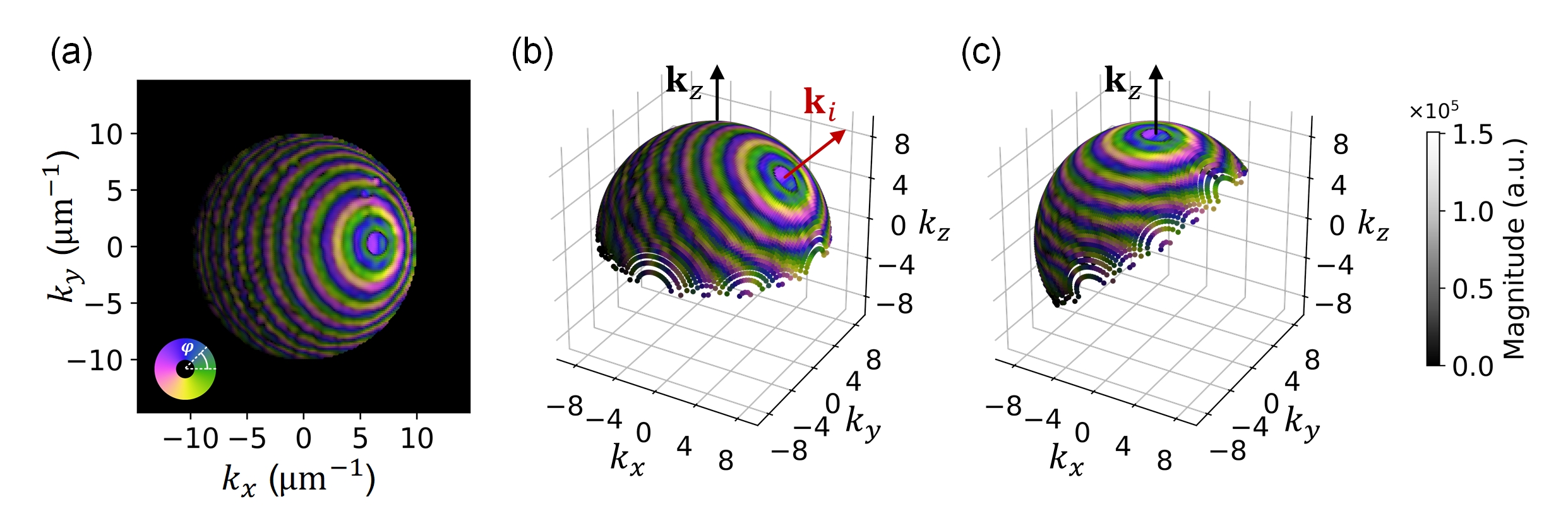}
    \caption{
        \textbf{ Measured co-polarized component of the complex scattering amplitude in $k$-space and angle space for a single incident vector $\mathbf{k}_i=(6.361,\, 0.326,\, 7.620)$~\textmu m$^{-1}$.} (a) Projection onto the plane spanned by $\mathbf{k}_x$ and $\mathbf{k}_y$. (b) Three-dimensional representation on the constant-$|\mathbf{k}|$ shell with the corresponding incident wave vector $\mathbf{k}_i$. (c) The same shell after rotation, so the scattering amplitude can be indexed by its angular coordinates. Complex plots: color encodes the phase and brightness encodes the magnitude \cite{kovesi2015good}. 
    }
    \label{fig:figure6}
\end{figure}

For a single measurement, the scattering amplitude vector $\mathbf{A}$ is associated with an incident wave vector $\mathbf{k}_i$. In off-axis holography, the scattering amplitudes are reconstructed on the plane spanned by $\hat{\mathbf{k}}_x$ and $\hat{\mathbf{k}}_y$ as shown in Fig. \ref{fig:figure6}(a), which corresponds to a 2D projection of the $|\mathbf{k}|=k_m$ sphere. When the illumination direction is tilted away from the optical axis, $\mathbf{k}_i$ rotates on the sphere and makes a polar angle $\theta_i$ with $\hat{\mathbf{k}}_z$, as shown in Fig. \ref{fig:figure6}(b), so the same scattering direction falls on different $k$-space locations during angle scanning.

To express scattering amplitude values measured with different illumination angles in the same spherical coordinates, the wave vectors $\{\mathbf{k}_n\}_{n=1}^{N}$ associated with $\mathbf{A}$ are rotated so that $\mathbf{k}_i$ is aligned with $\hat{\mathbf{k}}_z$.
The rotation is written as
\begin{equation}
\mathbf{k}'_n=\mathbf{R}(\hat{\mathbf{u}},\theta_i)\,\mathbf{k}_n\;,
\qquad
\hat{\mathbf{u}}=\frac{\mathbf{k}_i\times\hat{\mathbf{k}}_z}{\left\|\mathbf{k}_i\times\hat{\mathbf{k}}_z\right\|}\;,
\end{equation}
where $\mathbf{k}_n$ denotes one of the wave vectors, and $\mathbf{k}'_n$ is the same wave vector after rotation. The rotation matrix $\mathbf{R}(\hat{\mathbf{u}},\theta_i)$ is given by Rodrigues' formula \cite{cheng1989historical}, and $\hat{\mathbf{u}}$ is a unit vector that defines the rotation axis. With this incident alignment, as plotted in Fig. \ref{fig:figure6}(c), the rotated wave vectors $\{\mathbf{k}_n\}_{n=1}^{N}$ are converted to spherical coordinates and represented by the corresponding angle vectors $\{\mathbf{s}_n\}_{n=1}^{N}$ with $\mathbf{s}_n=(\theta_n,\phi_n)$ on the spherical shell.

By collecting the scattering amplitude values and corresponding angle vectors for all $N_{\mathrm{ill}}$ illumination angles, the measurement samples a broad set of scattering modes and covers the accessible ranges of $\theta$ and $\phi$. To combine all illumination angles, the field vectors are stacked into matrices,
\begin{equation}
\mathbf{T}_\mathrm{glass} =
\begin{bmatrix}
\mathbf{E}_\mathrm{glass}^{(1)} & \mathbf{E}_\mathrm{glass}^{(2)} & \cdots & \mathbf{E}_\mathrm{glass}^{(N_{\mathrm{ill}})}
\end{bmatrix}\in\mathbb{C}^{N\times N_{\mathrm{ill}}},\\
\;\;
\mathbf{T}_s =
\begin{bmatrix}
\mathbf{E}_s^{(1)} & \mathbf{E}_s^{(2)} & \cdots & \mathbf{E}_s^{(N_{\mathrm{ill}})}
\end{bmatrix}\in\mathbb{C}^{N\times N_{\mathrm{ill}}}.
\end{equation}
After rotating all measurements into the same incident reference ($\mathbf{k}_i = k_m\hat{\mathbf{k}}_z$), repeated observations of the same scattering direction occur among different illumination angles.
A shell average is then performed by averaging the scattering amplitude values that correspond to the same angle vector $\mathbf{s}$:
\begin{equation}
\bar{A}(\boldsymbol{\mathbf{s}})
=\frac{1}{N(\boldsymbol{\mathbf{s}})}\sum_{p\in\mathcal{P}(\boldsymbol{\mathbf{s}})} A^{(p)}(\mathbf{s}),
\end{equation}
where $\mathcal{P}(\mathbf{s})$ is the list of illumination angles that provide a measurement at direction $\mathbf{s}$, and $N(\mathbf{s})$ is the number of contributing measurements.
This produces a single complex scattering amplitude on the shell as a function of angle vector $\bar{A}(\mathbf{s})$, which can be directly compared across the full measured angular range.

\section{Experimental results}
The experimentally measured transmission and reflection matrices are then compared with the theoretical predictions based on the Mie scattering theory, enabling quantitative validation under the same illumination and polarization conditions as used in the optical setup. Both co-polarized and cross-polarized components of the scattering amplitude are evaluated.

\subsection{Forward scattering}
\begin{figure}[t]
    \centering
    \includegraphics[width=\linewidth]{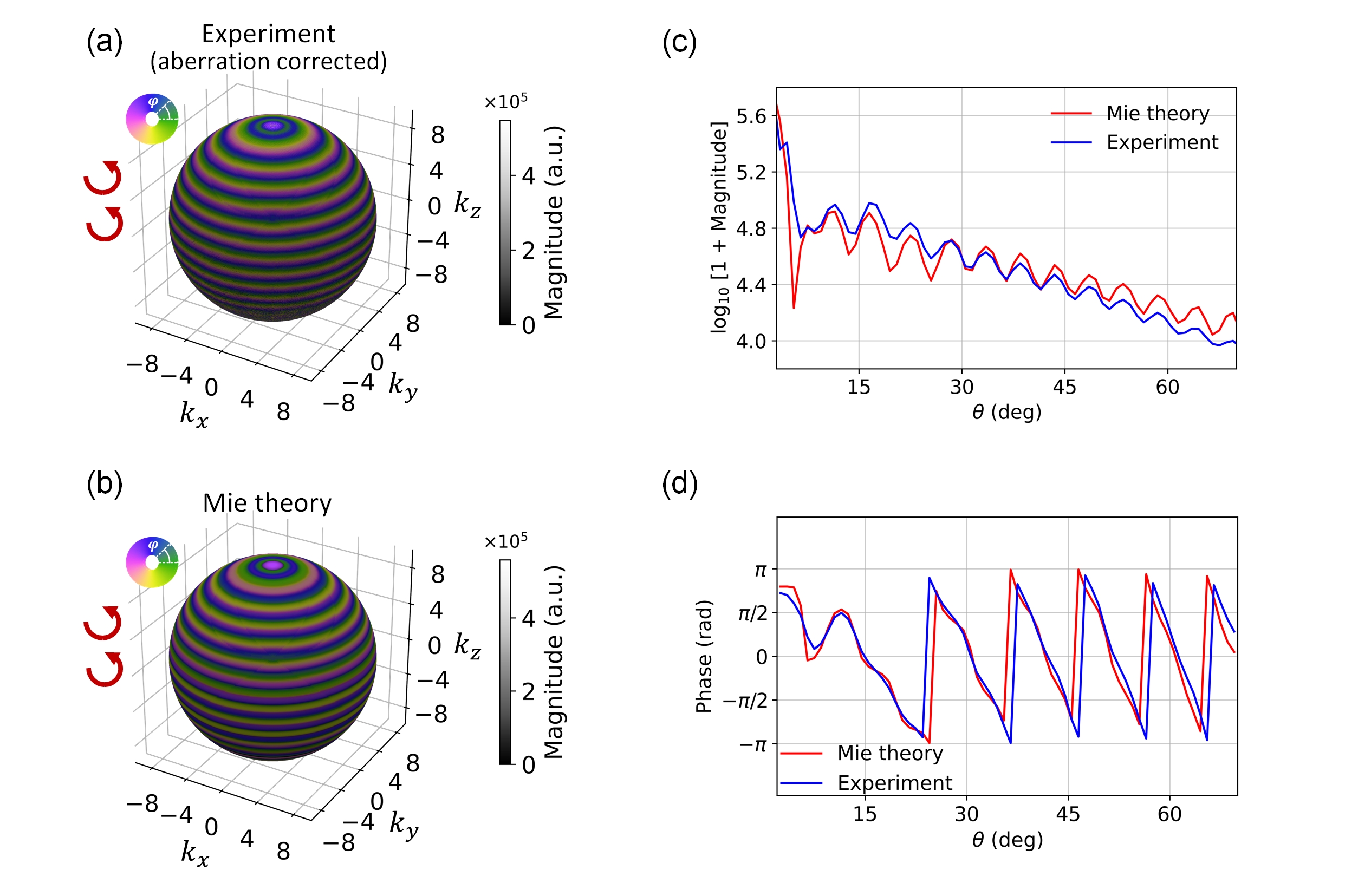}
    \caption{
        \textbf{Co-polarized scattering amplitude measured in the transmission experiment and compared with a fitted single-sphere Mie theory.} (a) Measured scattering amplitude on the transmission shell. (b) Simulated scattering amplitude from the fitted Mie theory. (c) Radially averaged magnitude over $\theta \in \left[3^\circ,\,70^\circ\right]$. (d) Radially averaged phase over $\theta \in \left[3^\circ,\,70^\circ\right]$. Complex plots: color encodes the phase and brightness encodes the magnitude.
    }
    \label{fig:figure7}
\end{figure}

Taking the dataset measured through the transmission path, the co-polarized scattering amplitude on the shell $\bar{A}(\mathbf{s})$ is shown in Fig. \ref{fig:figure7}(a), as a function of the scattering angle $\theta$ and the azimuthal angle $\phi$. As expected, the strongest response remains concentrated around the forward channel, and the magnitude gradually decreases toward larger scattering angles, indicating the dominance of low-angle scattering in the transmission geometry. The expected ring-like structure is also clearly visible, consistent with the fact that the single-sphere scattering amplitude is mainly a function of the scattering angle, so channels with similar $\theta$ form nearly concentric contours on the shell. There exist small deviations from perfect azimuthal symmetry, which arise from experimental factors such as optical noise, inaccurate $k$-space mapping, and finite angular sampling.

To link this measured angular pattern to physical parameters of the particle, we fit a single-sphere Mie model to the experimental scattering amplitude distribution. The experimentally observed values on this shell are assembled into a vector $\bar{\mathbf{A}}$ on the same set of angles as in Fig.~\ref{fig:figure7}(a). For a given set of parameters $\mathbf{x}$, the forward Mie model is evaluated at the corresponding scattering angles to generate the theoretical scattering amplitude vector $\mathbf{A}_\mathrm{Mie}$, and the parameters are obtained by minimizing a robust difference between the model and the data:
\begin{equation}
\mathbf{x}^\star
=
\min_{\mathbf{x}}\;\left\|\, g(\mathbf{x})\,\mathbf{A}_\mathrm{Mie}(\mathbf{x})-\bar{\mathbf{A}} \,\right\|_{1}\;,
\qquad \mathbf{x}=(n_p,\,a_r,\, z_f,\,k_\Delta)\;.
\end{equation}
Here, $\mathbf{x}$ collects the unknown sphere parameters: $n_p$ is the refractive index of the particle, $a_r$ is the sphere radius, $z_f$ describes the precise axial position of the focal plane in the particle coordinates, and $k_\Delta$ denotes the $k$-space pixel size in the holographic image. It is linked to the effective microscope magnification and sets the $k$-space grid, which in turn determines the angular sampling included in the fitting. The factor $g(\mathbf{x})$ is a scalar gain that accounts for an unknown overall energy scaling between experiment and theory, which enables the fitting to focus on the angular profile of the scattering. The function $\left\|\cdot\right\|_{1}$ denotes the Smooth–L1 loss, which is quadratic when the difference is small and becomes linear when the difference is large, reducing the impact of extreme values while remaining differentiable for gradient-based optimization (AdamW optimizer \cite{loshchilov2017decoupled}). The fitting yields $\mathbf{x}^\star$, with $n_p^\star\approx1.453$, $a_r^\star\approx4.057$~\textmu m, $z_f^\star\approx-0.045$~\textmu m, and $k_\Delta^\star\approx0.163$~\textmu$\mathrm{m}^{-1}$. The optimized parameters are inserted back into the forward Mie model to generate the simulated scattering amplitude shown in Fig.~\ref{fig:figure7}(b), where the propagation prefactor is compensated using the optimized axial position $z_f$. This simulation reproduces both the overall forward-dominated distribution and the main angular features.

To quantify the consistency, we radially average the shell data over the azimuthal angle $\phi$ to obtain 1D profiles as a function of scattering angle. The magnitude profiles in Fig. \ref{fig:figure7}(c) show that the experimental curve follows the Mie prediction, capturing the same Mie modes and the overall decay toward larger angles, with an RMSE of 0.149 between the two magnitude curves. The remaining mismatch is most noticeable at small scattering angles, where the measurement is close to the direct term and is therefore more sensitive to imperfect subtraction and normalization. At larger angles, the scattered signal becomes weak, which reduces the fringe contrast.
The corresponding phase profile in Fig. \ref{fig:figure7}(d) closely matches the Mie result over the full angular range, and remains in agreement even at large angles where the magnitude is reduced. Small differences are mainly attributed to the non-ideal sphere sample, since the particle is embedded on a glass substrate, as well as to inter-illumination variations in the fitted axial position $z_f$.

\begin{figure}[t]
    \centering
    \includegraphics[width=\linewidth]{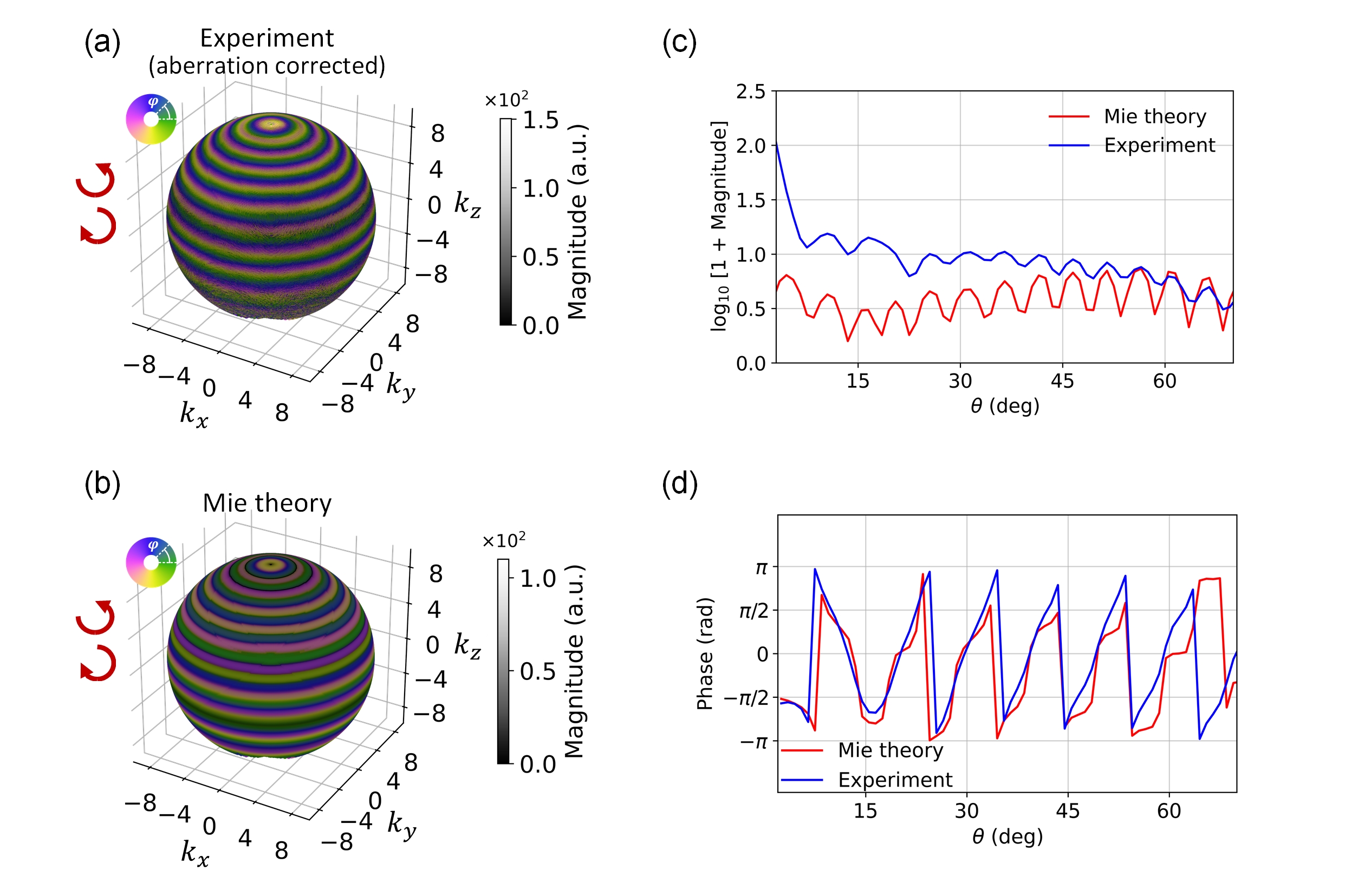}
    \caption{
        \textbf{Cross-polarized scattering amplitude in the transmission experiment and comparison with Mie-theory prediction.} (a) Measured scattering amplitude on the transmission shell. (b) Simulated scattering amplitude calculated using the parameters obtained from the co-polarized fitting. (c) Radially averaged magnitude over $\theta \in \left[3^\circ,\,70^\circ\right]$. (d) Radially averaged phase over $\theta \in \left[3^\circ,\,70^\circ\right]$. Complex plots: color encodes the phase and brightness encodes the magnitude.
    }
    \label{fig:figure8}
\end{figure}

A similar analysis is carried out for the cross-polarized component. The experimentally measured cross-polarized scattering amplitude on the transmission shell is shown in Fig.~\ref{fig:figure8}(a), where the magnitude is generally weaker than in the co-polarized channel but still exhibits a clear angular structure. For comparison with Mie theory, we keep the sphere parameters fixed, which are calculated from the fitted co-polarized data, and the simulation is shown in Fig.~\ref{fig:figure8}(b).

Similarly, the cross-polarized scattering amplitude shell is averaged into a 1D profile by azimuthal averaging. Compared with the co-polarized channel, the cross-polarized magnitude is much weaker, so the fine Mie-mode oscillations are harder to resolve. In particular, the small-angle region has much lower contrast, so the fine rings that appear clearly in the co-polarized profile are barely visible in the cross-polarized measurement. At intermediate and larger angles, the stronger Mie-mode features remain evident, and the simulation reproduces both the overall decay and the overall oscillation pattern. The reduced low-angle mode visibility is mainly due to the weak cross-polarized signal and the stronger influence of the noise floor, and it is further affected by co-polarized leakage from imperfect wave-plate alignment. Despite these limitations, the magnitude curves in Fig.~\ref{fig:figure8}(c) still match qualitatively. The experimental phase curve in Fig.~\ref{fig:figure8}(d) agrees surprisingly well with the Mie result, even with the lower signal-to-noise ratio in the cross-polarized channel.

\subsection{Reflection matrix}

Compared with forward scattering, the backward scattering from a single dielectric sphere of this size parameter is inherently weaker and occupies a smaller effective angular range in the condenser pupil, making the measurement more sensitive to noise and system errors. In addition, reflection measurements are influenced by the air-glass interface, because the multiple reflections between the sphere and the interface are coherently added to the particle response, which introduces additional modulation in the measured field. The reconstructed co-polarized component of the back-scattered field is right-handed, as the reflection process inherently flips the incident circular polarization. Using the sphere parameters obtained from the forward-scattering fitting, the Mie-theory prediction approximated by Eq. (\ref{eq:four_terms_from_two_sums}) is evaluated for the corresponding backward directions and compared with the measured scattered fields. This comparison provides an independent consistency check of the same sphere parameters in a different scattering geometry.

\begin{figure}[t]
    \centering
    \includegraphics[width=\linewidth]{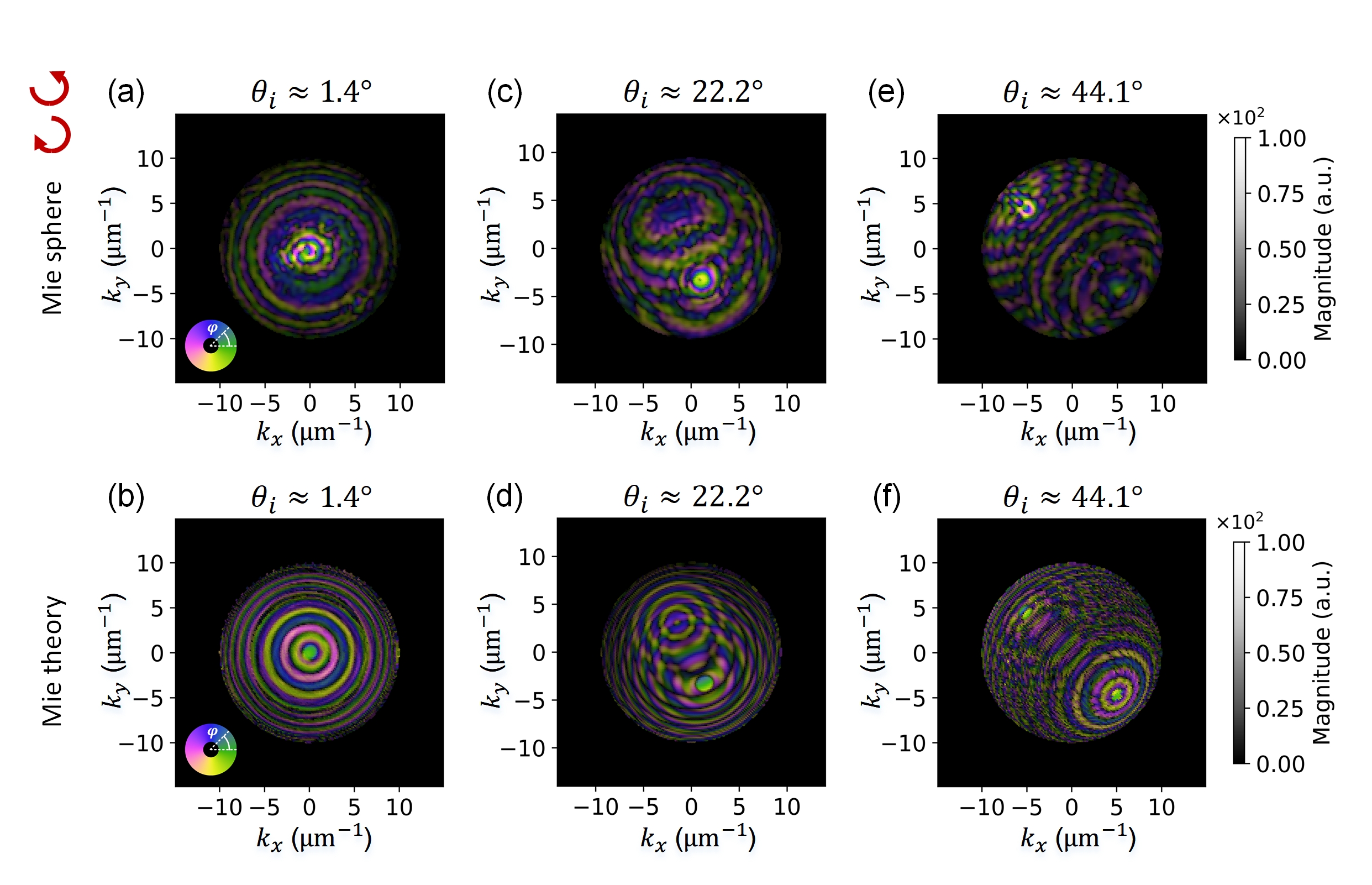}
    \caption{
        \textbf{ Comparison of experimentally reconstructed and simulated scattered fields in reflection for three illumination angles.} (a), (c), and (e) show the measured complex field on the pupil plane for $\theta_i=1.4^\circ$, $22.2^\circ$, and $44.1^\circ$, respectively. (d), (d), and (f) show the corresponding Mie-theory predictions evaluated under the same illumination geometry. Complex plots: color encodes the phase and brightness encodes the magnitude. 
    }
    \label{fig:figure9}
\end{figure}

The experimentally reconstructed complex scattered field from the reflection path is shown in Figs.~\ref{fig:figure9}(a), (c), and (e) for three representative illumination angles, $\theta_i\approx1.4^\circ$, $22.2^\circ$, and $44.1^\circ$. The corresponding simulations in Figs.~\ref{fig:figure9}(b), (d), and (f) are evaluated under the same illumination geometry. Good qualitative agreement is observed despite noise.

With nearly on-axis illumination $\theta_i\approx1.4^\circ$, the measured scattered field exhibits a predominantly concentric-rings structure centered close to the pupil origin. This pattern mainly reflects the azimuthally symmetric reflection response of the sphere, with additional azimuthally symmetric modulation from the terms corresponding to forward scattering, preceded and followed by reflection at the glass surface. The simulation reproduces the same global symmetry and the dominant ring positions, indicating that the main scattering phase accumulation is described by the model.

At the intermediate angle $\theta_i\approx22.2^\circ$, both the measurement and the simulation become clearly asymmetric across the pupil. The rings are no longer centered, and the dominant fringes shift laterally and vary with azimuthal angle because of the tilted illumination. In the lower-right region, the bright concentric rings contain a strong interface-driven contribution, dominated by the terms G, G-MieFw, MieFw-G, and a substantial part of MieFw-G-MieFw. The Mie-mode content in these rings is therefore mainly carried by the sphere transmission contribution that is coupled into the backward measurement by substrate reflection. By contrast, the weaker secondary rings toward the upper-left are mainly associated with the direct sphere response MieBw. The simulation captures the overall fringe orientation and curvature, while small local differences are caused by other reflection aberrations, finite angle sampling, and higher-order multiple reflections beyond the first five terms.

At the larger angle $\theta_i\approx44.1^\circ$, the pupil field becomes highly structured with dense oscillations. The fringes in the measurement indicate strong interference between multiple angular components in the reflected field, and the simulation reproduces a similar increase in fringe density with larger incident angle. Here, the fringes in the upper-left are dominated by the interface-driven terms  G, G-MieFw, MieFw-G, with only a weak contribution from MieFw-G-MieFw under stronger tilt. The lower-left region is mainly associated with the direct sphere response MieBw. The remaining differences are more evident at large angles because the signal-to-noise ratio drops near the pupil edge and the measurement becomes more sensitive to polarization-dependent Fresnel reflection at the air-glass interface, which adds extra phase and amplitude modulation.

Overall, Fig.~\ref{fig:figure9} shows that the back-scattered field changes strongly with illumination angle, and that the forward Mie model captures the main scattering features on the condenser pupil over a broad angular range. This result provides experimental support for the reflection series introduced in Eq. (\ref{eq:four_terms_from_two_sums}). The comparison also indicates that substrate coupling and optical imperfections have a more significant impact on the reflection matrix, compared with the transmission matrix.

\section{Conclusion}
In this work, we measured the polarization-complete transmission matrix of a single dielectric sphere and used it to recover its angle-dependent scattering response, which is well described by Mie theory. In the co-polarized transmission data, the scattering is strongly forward-directed and shows concentric Mie-mode features in angle space after aberration correction. The parameters fitted from this channel remain predictive when applied to the cross-polarized data and to the reflection measurements. For reflection, matching the measured fields requires including coherent scattering contributions arising from the glass substrate, and the consistency in the comparison supports both the forward model and the sphere-glass coupling mechanism.

These results show that a measured transmission matrix in the angular basis can be converted into the scattering amplitude that is directly linked to physical particle parameters. Once the mode coupling is expressed as an angular scattering amplitude, the size and refractive index of a spherical or quasi-spherical scatterer can be determined from the measurements using a forward-model fitting, and then validated across polarization channels and measurement geometries. Importantly, correcting pupil aberrations and expressing the transmission matrix in angle space make the three-dimensional scattering process interpretable at the field level. This approach also provides a practical basis for quantitative analysis of more complex scatterers, with further improvements expected from better polarization calibration, calculation with quasi-spherical clusters, and more complete forward models.

\section*{Acknowledgements}
We thank A. van Blaaderen, R. Yang, and J. Bückmann for sample preparation and experimental suggestions, W. Vos, L. Filion, A. Imhof, P. van der Straten, D. van Oosten,  N. Alferink, and B. Verreusel for insightful discussions, and J.B. Aans, A. Driessen, P. van den Beld, and D. Killian for technical support.

\paragraph{Funding information}
This work was supported by the Netherlands Organization for Scientific Research NWO (ENW grant number OCENW.GROOT.2019.071).

\begin{appendix}
\numberwithin{equation}{section}
\section{Appendix: Mie scattering theory}
\label{appendix:mie_theory}
For a monochromatic electromagnetic field in a linear, isotropic, homogeneous medium, the field components satisfy the Helmholtz equation \cite{born2013principles}. Introducing a scalar function $\psi$ that satisfies the scalar Helmholtz equation in spherical coordinates $(r,\theta,\phi)$,
\begin{equation}
\frac{1}{r^{2}}\frac{\partial}{\partial r}\!\left(r^{2}\frac{\partial \psi}{\partial r}\right)
+\frac{1}{r^{2}\sin\theta}\frac{\partial}{\partial \theta}\!\left(\sin\theta\frac{\partial \psi}{\partial \theta}\right)
+\frac{1}{r^{2}\sin^{2}\theta}\frac{\partial^{2}\psi}{\partial \phi^{2}}
+k_m^{2}\psi=0\;,
\label{eq:scalar_wave_spherical}
\end{equation}
with $k_m^{2}=\omega^{2}\epsilon\mu$, where $\epsilon$ is the permittivity and $\mu$ the permeability, a convenient set of divergence-free vector solutions is obtained from a scalar solution $\psi$ of the Helmholtz equation by defining two vector wave functions
\begin{equation}
\mathbf{M} = \nabla \times \left(\mathbf{r}\,\psi\right)\;,
\qquad
\mathbf{N} = \frac{1}{k_m}\,\nabla \times \mathbf{M}\;,
\label{eq:MN_def}
\end{equation}
where $\mathbf{r}$ is the position vector and $k_m$ is the wave number of the surrounding medium. The pair $(\mathbf{M},\mathbf{N})$ named as vector spherical harmonics form the basis used to expand the incident, internal, and scattered fields of a sphere.

To characterize the scattering response of a single, well-defined sphere, we consider plane-wave illumination with an incident wave vector $\mathbf{k}_i$ aligned with the optical axis. The scattered field is observed in direction $\mathbf{k}_s$, which specifies the propagation direction of a plane-wave component from the scattered field. We introduce the corresponding unit vectors $\hat{\mathbf{k}}_i$ and $\hat{\mathbf{k}}_s$, where $\hat{\mathbf{k}}_i$ is temporarily placed along the $z$ axis for the convenience of description. The scattering plane is then defined as the plane spanned by $\hat{\mathbf{k}}_i$ and $\hat{\mathbf{k}}_s$. Based on this plane, the s- and p-polarization bases are defined. The s-polarization direction is perpendicular
to the scattering plane:
\begin{equation}
\hat{\mathbf{e}}_{\perp}
=
\frac{\hat{\mathbf{k}}_i\times \hat{\mathbf{k}}_s}
{\lVert \hat{\mathbf{k}}_i\times \hat{\mathbf{k}}_s\rVert}\;,
\label{eq:e_s_def}
\end{equation}
and the p-polarization directions lie in the scattering plane and are orthogonal to the propagation direction:
\begin{equation}
\hat{\mathbf{e}}_{{\parallel}i}
=
\hat{\mathbf{k}}_i \times \hat{\mathbf{e}}_{\perp}\;,
\qquad
\hat{\mathbf{e}}_{{\parallel}s}
=
\hat{\mathbf{k}}_s \times \hat{\mathbf{e}}_{\perp}\;.
\label{eq:e_p_def}
\end{equation}
Here, $\perp$ denotes the field component of s-polarization (perpendicular),
and $\parallel$ denotes the component of p-polarization (parallel). Given the incident electric field $E_{i}$ and the scattered electric field $E_{s}$, it accordingly yields the s- and p-components of the fields:
\begin{equation}
E_{\perp i}=\mathbf{E}_i\cdot \hat{\mathbf{e}}_{\perp}\;,\quad
E_{\parallel i}=\mathbf{E}_i\cdot \hat{\mathbf{e}}_{{\parallel}i}\;,\quad
E_{\perp s}=\mathbf{E}_s\cdot \hat{\mathbf{e}}_{\perp}\;,\quad
E_{\parallel s}=\mathbf{E}_s\cdot \hat{\mathbf{e}}_{{\parallel}s}\;.
\label{eq:field_components_def}
\end{equation}
The observation direction $\hat{\mathbf{k}}_s$ can be parameterized by the scattering angle $\theta$ and the azimuthal angle $\phi$ with respect to the optical axis. The scattering angle $\theta$ is the polar angle of the scattered wave vector
$\mathbf{k}_s$ with respect to the $z$ direction:
\begin{equation}
\theta \;=\; \arccos\!\left(\frac{\mathbf{k}_s\cdot \hat{\mathbf{k}}_z}{\lVert \mathbf{k}_s\rVert}\right)\;,
\label{eq:theta_definition_kk}
\end{equation}
where $\hat{\mathbf{k}}_z$ is the unit vector pointing along the $k_z$ axis. The azimuthal angle $\phi$ is defined as the angle of the projection of $\hat{\mathbf{k}}_s$ onto the transverse $(k_x,k_y)$ plane, measured from the $k_x$ axis:
\begin{equation}
\phi=\operatorname{atan2}\!\left(\hat{\mathbf{k}}_s\cdot \hat{\mathbf{k}}_y,\,\hat{\mathbf{k}}_s\cdot \hat{\mathbf{k}}_x\right)\;,
\label{eq:phi_definition}
\end{equation}
where $\hat{\mathbf{k}}_x$ and $\hat{\mathbf{k}}_y$ are unit vectors along the $k_x$ and $k_y$ axes.

In the far field, the relation between the incident and scattered electric-field components can be written in terms of the amplitude scattering matrix
\begin{equation}
\begin{pmatrix}
E_{\parallel s} \\
E_{\perp s}
\end{pmatrix}
=
\frac{e^{ik(r-z)}}{-ikr}
\begin{pmatrix}
S_{1} & S_{3} \\
S_{4} & S_{2}
\end{pmatrix}
\begin{pmatrix}
E_{\parallel i} \\
E_{\perp i}
\end{pmatrix}\;.
\label{eq:amp_scattering_matrix_general}
\end{equation}
The exponential prefactor accounts for the spherical-wave decay and phase accumulation in the far field. The coefficients $S_{1}$, $S_2$, $S_3$ ,and $S_{4}$ depend on the scattering angle $\theta$ and the azimuthal angle $\phi$. For a homogeneous sphere, rotational symmetry removes any dependence on $\phi$, and polarization mixing terms vanish, such that $S_{3}=S_{4}=0$, and the remaining elements are expressed as the uniformly convergent series:
\begin{align}
S_{1}(\theta)
&=
\sum_{n=1}^{\infty}\frac{2n+1}{n(n+1)}
\left[a_{n}\,\pi_{n}(\theta)+b_{n}\,\tau_{n}(\theta)\right]\;,
\label{eq:S1}\\
S_{2}(\theta)
&=
\sum_{n=1}^{\infty}\frac{2n+1}{n(n+1)}
\left[a_{n}\,\tau_{n}(\theta)+b_{n}\,\pi_{n}(\theta)\right]\;,
\label{eq:S2}
\end{align}
where $a_{n}$ and $b_{n}$ are the external field coefficients, which are also named as Mie coefficients. They depend on the size parameter and the complex refractive index of the particle. The angular functions $\pi_n$ and $\tau_n$ are defined in terms of associated Legendre functions $P_n^1$:
\begin{equation}
\pi_{n}(\theta)=\frac{P_{n}^{1}(\cos\theta)}{\sin\theta}\;,
\qquad
\tau_{n}(\theta)=\frac{d}{d\theta}P_{n}^{1}(\cos\theta)\;,
\label{eq:pi_tau_def}
\end{equation}
which can be computed efficiently by upward recurrence. Let $x$ denote the size parameter, and the relative refractive index is $m$, as defined by Eq. (\ref{eq:size_parameter}) in Section \ref{subsec:mie_theory}. Using Riccati-Bessel functions $\psi_n(\rho)=\rho\,j_n(\rho)$ and $\xi_n(\rho)=\rho\,h_n^{(1)}(\rho)$, the Mie coefficients can be written as
\begin{align}
a_n &=
\frac{ m\psi_n(mx)\psi_n'(x)-\psi_n(x)\psi_n'(mx)}
{ m\psi_n(mx)\xi_n'(x)-\xi_n(x)\psi_n'(mx)}\;,
\label{eq:an}\\
b_n &=
\frac{ \psi_n(mx)\psi_n'(x)-m\psi_n(x)\psi_n'(mx)}
{ \psi_n(mx)\xi_n'(x)-m\xi_n(x)\psi_n'(mx)}\;.
\label{eq:bn}
\end{align}

\section{Appendix: Off-axis holographic microscopy}
\label{appendix:Off-axis holography}
\begin{figure}[t]
    \centering
    \includegraphics[width=\linewidth]{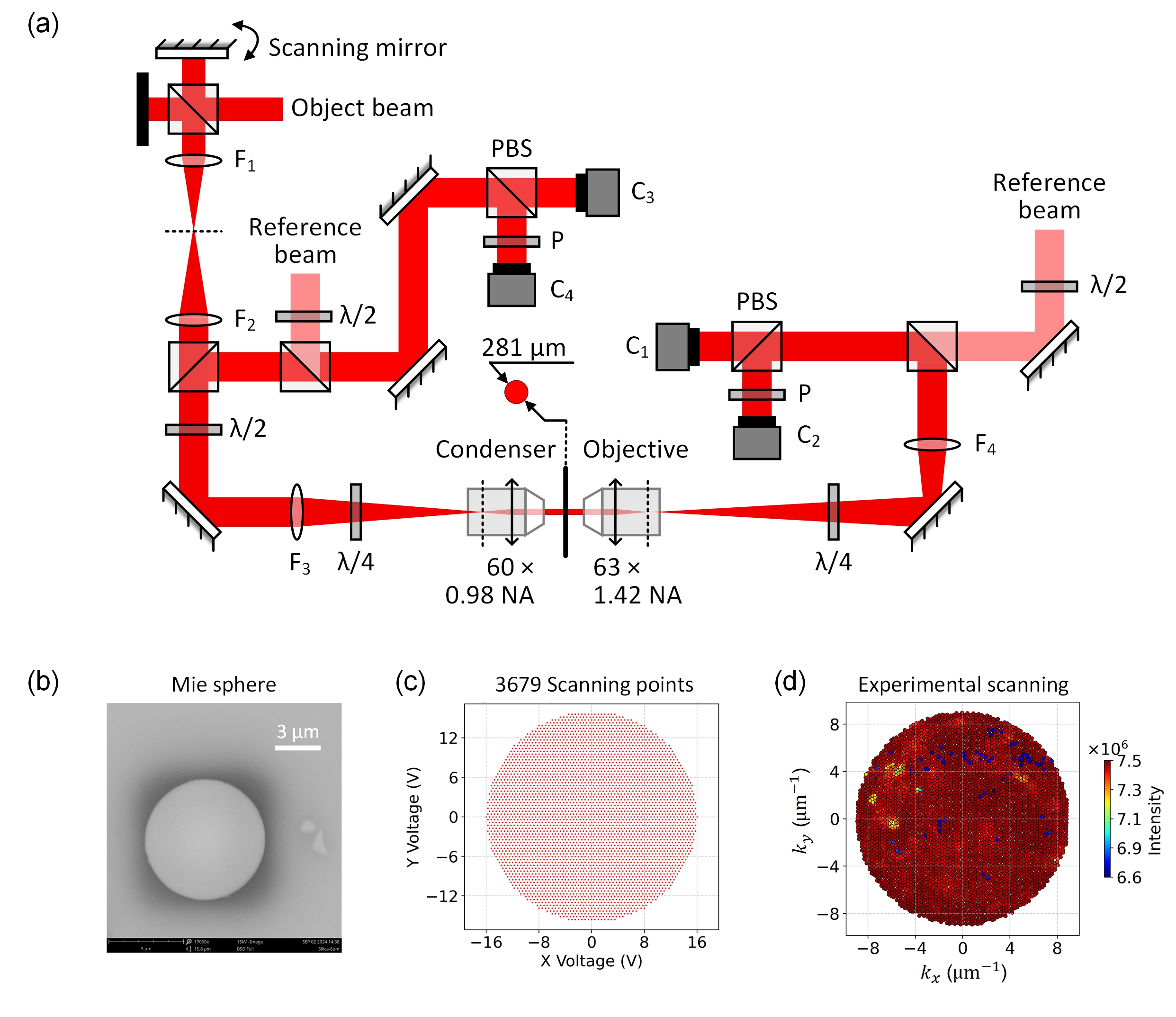}
    \caption{
        \textbf{Experimental setup and illumination scanning.} (a) Off-axis holographic microscopy. (b) Bright-field image of the microsphere from scanning electron microscopy (SEM). The particle has an approximate diameter of 8~\textmu m and a refractive index of $\sim$1.45 (supplier Bangs Laboratories, Fisher, USA). (c) Hexagonal-spiral scanning pattern used to sample the illumination pupil. (d) Total field intensity measured over the scanned pupil.
    }
    \label{fig:figure2}
\end{figure}

The experimental setup is an interferometric microscope configured based on off-axis digital holography \cite{leith1962reconstructed, takeda1982fourier} with angle-scanned illumination, as illustrated in Fig. \ref{fig:figure2}(a). The light from a He-Ne laser source (wavelength $= 632.8$ nm) is polarized and split into an object beam and a reference beam. 

In the path of the object beam, a scanning mirror steers the illumination direction and is placed in a plane conjugate to the back focal plane of the lens $F_2$, which is also the front focal plane of the lens $F_3$. As a result, mirror rotation is converted into a tilt of the plane wave at the sample while largely maintaining the beam position. The object beam is then guided into the microscope and illuminates the particle through a condenser objective (60$\times$, NA~0.98). The forward-scattered field is collected by an imaging objective (63$\times$, NA~1.42), while the back-scattered field from the particle and the air-glass interface is collected by the condenser objective and routed into a reflection path. For both paths, lenses $F_{3}$ and $F_{4}$ form an imaging layout that places the camera plane in conjugation with the particle plane. This ensures that the detector is optically focused on the particle plane, which is the $z=0$ plane of the particle-centered coordinate system defined in Section \ref{subsec:mie_theory}.

The reference beam propagates along a separate path and passes through a half-wave plate ($\lambda/2$, HWP) to match the polarization state of the selected object-field component. To realize off-axis holography, the reference beam is introduced with a small angular tilt relative to the object beam at the camera plane. For both transmission and reflection measurements, the corresponding object field is directed to the interferometer and recombined with the reference beam at a polarizing beam splitter (PBS), producing off-axis holograms for complex-field reconstruction.

The illumination polarization onto the particle is prepared with a HWP followed by a quarter-wave plate ($\lambda/4$, QWP). The HWP is set to $45^\circ$ with respect to the experimental $x$--$y$ axes to rotate the input linear polarization to the required orientation, and the subsequent QWP converts this linear state into circular polarization at the sample.

In the transmission path, circular components are analyzed using a second QWP placed before a polarizing beam splitter (PBS). With the chosen orientation, this QWP maps the left- and right-circular components onto orthogonal linear polarizations: the left-circular component is converted to $y$ polarization and the right-circular component to $x$ polarization. The PBS then separates the two components into different output ports, allowing the complex fields associated with $E_L$ and $E_R$ to be measured independently. A clean-up polarizer (P) is placed on the reflection-side output of the PBS to suppress leakage from the orthogonal component. In the reflection path, the back-scattered field retraces the illumination optics and therefore passes through the first QWP again. This second pass performs the same circular-to-linear conversion. The PBS subsequently splits the $x$- and $y$-polarized components into two spatially separated ports for independent detection. 

The illumination angles are adjusted by driving the two-axis scanning mirror along a hexagonal spiral trajectory, starting near the pupil center and expanding to cover the entire pupil. This hexagonal spiral provides an approximately uniform sampling distance on the illumination pupil, as is plotted in Fig. \ref{fig:figure2}(c), so illumination wave vectors are sampled at nearly constant spacing. To improve the precision of the scanning mirror, we smoothly scan the driving voltage between adjacent scanning positions. Based on a measurement on an empty glass substrate, Fig.~\ref{fig:figure2}(d) shows that the total recorded field intensity varies by less than \(10\%\) across the pupil.

For a single arbitrary polarization channel, one recorded hologram can be written as:
\begin{equation}
I = \left|E_{\mathrm{obj}} + E_{\mathrm{ref}}\right|^{2}
       = |E_{\mathrm{obj}}|^{2} + |E_{\mathrm{ref}}|^{2}
       + E_{\mathrm{obj}}E_{\mathrm{ref}}^{*} + E_{\mathrm{obj}}^{*}E_{\mathrm{ref}}\;,
\end{equation}
where $E_{\mathrm{obj}}$ and $E_{\mathrm{ref}}$ denote the complex object and reference fields at the detector, respectively. The off-axis spatial frequency introduced by the tilted reference beam shifts the terms $E_{\mathrm{obj}}E_{\mathrm{ref}}^{*}$ and $E_{\mathrm{obj}}^{*}E_{\mathrm{ref}}$ to opposite sidebands in the Fourier domain, allowing the isolation of a single term by spatial filtering. The complex object field is then recovered by inverse Fourier transform. Angle scanning of the illumination yields a set of off-axis holograms at different incident wave vectors for subsequent transmission matrix analysis.
\end{appendix}




\bibliography{SciPost_Example_BiBTeX_File.bib}

\end{document}